\documentclass[aps,prd,preprint,superscriptaddress,showpacs]{revtex4}

\usepackage{amsfonts}
\usepackage{mathrsfs}
\usepackage{amssymb}
\usepackage{amsmath}
\usepackage{graphicx,epsfig}

\def\bwt{\begin{widetext}}

\def\ewt{\end{widetext}}

\def\be{\begin{equation}}

\def\ee{\end{equation}}

\def\bea{\begin{eqnarray}}

\def\eea{\end{eqnarray}}

\def\bean{\begin{eqnarray*}}

\def\eean{\end{eqnarray*}}

\def\bary{\begin{array}}

\def\eary{\end{array}}

\def\bit{\begin{itemize}}

\def\eit{\end{itemize}}

\def\su5u1{SU(5) \times U(1)}

\def\fsu5u1{SU(5) \times U(1)'}

\def\so10{SO(10)}

\def\sq20{SO(10) \times SO(10)}

\usepackage{amssymb}

\begin{document}

\title{Background Dependent Lorentz Violation: Natural Solutions to
 the Theoretical Challenges of the OPERA Experiment}

\author{Tianjun Li}

\affiliation{ Key Laboratory of Frontiers in Theoretical Physics,
  Institute of Theoretical Physics, Chinese Academy of Sciences,
  Beijing 100190, P. R. China}

\affiliation{George P. and Cynthia W. Mitchell Institute for
Fundamental Physics and Astronomy, Texas A$\&$M University, 
College Station, TX 77843, USA }

\author{Dimitri V. Nanopoulos}

\affiliation{George P. and Cynthia W. Mitchell Institute for
Fundamental Physics and Astronomy, Texas A$\&$M University, 
College Station, TX 77843, USA }

\affiliation{Astroparticle Physics Group,
Houston Advanced Research Center (HARC),
Mitchell Campus, Woodlands, TX 77381, USA}

\affiliation{Academy of Athens, Division of Natural Sciences,
 28 Panepistimiou Avenue, Athens 10679, Greece }



\begin{abstract}

To explain both the OPERA experiment and all the known 
phenomenological constraints/observations on Lorentz 
violation, the Background Dependent Lorentz Violation 
(BDLV) has been proposed. We study the BDLV in 
a model independent way, and conjecture that {\it there 
may exist a ``Dream Special Relativity Theory'', where all 
the Standard Model (SM) particles can be subluminal 
due to the background effects}. Assuming that the Lorentz 
violation on the Earth is much larger than those on the 
interstellar scale, we automatically escape all the 
astrophysical constraints on Lorentz violation. For the 
BDLV from the effective field theory,  we present a simple 
model and discuss the possible solutions to the 
theoretical challenges of the OPERA experiment
such as the Bremsstrahlung effects for muon neutrinos 
and the pion decays. Also, we address the
Lorentz violation constraints from the LEP and KamLAMD 
experiments. For the BDLV from the Type IIB string 
theory with D3-branes and D7-branes,
we point out that {\it the D3-branes are flavour blind, and
 all the SM particles 
are the conventional particles as in the traditional
SM when they do not interact with the D3-branes}. Thus, 
we not only can naturally avoid all the known
phenomenological constraints on Lorentz violation, but 
also can naturally explain all the theoretical challenges.  
Interestingly, the energy dependent photon 
velocities may be tested at the experiments.

\end{abstract}

\pacs{11.10.Kk, 11.25.Mj, 11.25.-w, 12.60.Jv}

\preprint{ACT-17-11, MIFPA-11-47}

\maketitle

\section{Introduction}

The OPERA neutrino experiment at the underground Gran Sasso Laboratory (LNGS)
has recently determined the muon neutrino ($\nu_\mu$) velocity with high
accuracy through the measurement of the flight time and the distance (730 km)
between the source of the CNGS neutrino beam at CERN (CERN Neutrino beam 
to Gran Sasso) and the OPERA detector at the LNGS~\cite{opera}. The mean neutrino energy
is 17 GeV. Very surprisingly, the OPERA experiment found that neutrinos arrived
earlier than expected from luminal speed by a time interval
\begin{eqnarray}
\delta t = \left( 60.7 \pm 6.9_{\rm stat} \pm 7.4_{\rm syst} \right) {\rm ns}~.~\,
\end{eqnarray}
This implies a superluminal propagation velocity for neutrinos by a relative amount
\begin{eqnarray}
\delta v_\nu ~\equiv~  {{v_{\nu} -c}\over c} ~=~
\left( 2.48 \pm 0.28_{\rm stat} \pm 0.30_{\rm syst} \right) \times 10^{-5}~~~~~~{\rm (OPERA)},
\label{eq:opr}
\end{eqnarray}
where $c$ is the speed of light in the vacuum. Moreover, the 
  neutrino energy dependence for $\delta t$ was studied as well. For the neutrino
mean energies 13.9 GeV and 42.9 GeV, the experimental values of the
associated early arrival times are respectively
\begin{eqnarray}
\delta t_1 &=& \left( 53.1 \pm 18.8_{\rm stat} \pm 7.4_{\rm syst} \right) {\rm ns}~~~~~
{\rm for} ~{\langle E \rangle}_{\nu} = 13.9~{\rm GeV}~,~ \\
\delta t_2 &=& \left( 67.1 \pm 18.2_{\rm stat} \pm 7.4_{\rm syst} \right) {\rm ns}~~~~~
{\rm for} ~{\langle E \rangle}_{\nu} = 42.9~{\rm GeV}~.~ 
\end{eqnarray}
Thus, there is no significant
dependence for $\delta t$ on neutrino energies.

Interestingly, the OPERA results are compatible with
the MINOS results~\cite{minos}.  Although not statistically significant, the MINOS 
Collaboration has found~\cite{minos}
\begin{eqnarray}
\delta v_\nu =  \left( 5.1 \pm 2.9 \right) \times 10^{-5}~~~~~~{\rm (MINOS)}~,~
 \end{eqnarray}
for muon neutrino with a spectrum peaking at about 3 GeV, and a tail
extending above 100 GeV. Moreover, the earlier short-baseline experiments have set
the upper bounds on $|\delta v_\nu |$ around $4\times 10^{-5}$ in the energy range 
from 30 GeV to 200 GeV~\cite{old}. 
Of course, the technical issues in the OPERA 
experiment such as pulse modelling, timing and distance measurement deserve 
further experimental scrutiny. Other
experiments like MINOS and T2K are also needed to to do independent measurements
for further confirmation due to neutrino oscillations.
From theoretical point of view, many groups 
have already studied the possible solutions 
or pointed out the challenges to the OPERA anomaly~\cite{Cacciapaglia:2011ax,
AmelinoCamelia:2011dx, Giudice:2011mm, Dvali:2011mn, Mann:2011rd, Drago:2011ua,
Li:2011ue, Pfeifer:2011ve, Lingli:2011yn, 
Alexandre:2011bu, Cohen:2011hx, GonzalezMestres:2011jc, Matone:2011jd, 
Ciuffoli:2011ji, Bi:2011nd, Wang:2011sz, Cowsik:2011wv, Li:2011zm, 
AmelinoCamelia:2011bz, Moffat:2011ue, Faraggi:2011en}. For an early similar study,
see Ref.~\cite{Ellis:2008fc}.

The major challenges to the OPERA experimental results are the following: 
(1) Bremsstrahlung
effects~\cite{Cohen:2011hx}. The superluminal muon neutrinos with $\delta v_{\nu}$ given in 
Eq.~(\ref{eq:opr}) would lose energy rapidly via Cherenkov-like processes
on their ways from CERN to LNGS, and the most important process is
 $\nu_{\mu} \to \nu_{\mu} + e^+ + e^-$. Thus, the OPERA experiment can not observe
the muon neutrinos with energy in excess of 12.5 GeV~\cite{Cohen:2011hx}; 
(2) Pion decays~\cite{GonzalezMestres:2011jc, Bi:2011nd, Cowsik:2011wv}.
The superluminal muon neutrinos with $\delta v_{\nu}$  in  Eq.~(\ref{eq:opr}) can
not have energy larger than about 10 GeV from pion decays, for example,
$\pi^+ \to \mu^+ \nu_{\mu}$ and $\mu \to \nu_{\mu} + e + {\bar \nu}_e$~\cite{Bi:2011nd}. 
One solution to these challenges is that
these anomalous processes are forbidden if the Lorentz symmetry is deformed,
preserving the relativity of inertial frames~\cite{AmelinoCamelia:2011bz}. 
These deformations add non-linear
terms to the energy-momentum relations, conservation laws, and Lorentz
transformations in a way which is consistent with the relativity of inertial
observers. However, the studied model is a toy model, which 
can not explain the OPERA results~\cite{AmelinoCamelia:2011bz}.

To explain both the OPERA experiment and all the known phenomenological
constraints/observations on Lorentz violation, 
the Background Dependent Lorentz Violation (BDLV) has
been proposed by considering the spin-2, spin-1, spin-0 particles, and 
Type IIB string theory, respectively in 
Refs.~\cite{Dvali:2011mn, Alexandre:2011bu, Ciuffoli:2011ji, Li:2011zm}.
In this paper, we briefly review the relevant phenomenological constraints
and observations on Lorentz violation~\cite{imb, baks, kam, Albert:2007qk,
Aharonian:2008kz, Abdo:2009pg, Abdo:2009zza, :2009zq, Abbasi:2010ie, crab,
ems, crab2, bou2,  bou3, Altmu, bou1, :2008ee, review}, 
and we study the BDLV in a  model independent way. 
In particular, we conjecture that {\it there may exist a ``Dream Special Relativity 
Theory'', where all the Standard Model (SM) particles 
can be subluminal due to the background effects}.
We also suggest that the OPERA experiment can measure the 
velocities of the muon neutrinos with energies around a few GeV, which
may test whether the muon neutrino velocities depend on their energies or not.
Assuming that the Lorentz violation on
the Earth is much larger than those on the interstellar scale, we automatically
escape all the astrophysical constraints on Lorentz violation.
To explain the OPERA results
in the effective field theory approach,
we present a simple model with a triplet Higgs field 
as in the Type II seesaw mechanism~\cite{GRmodel, triplet1}, 
where we introduce the non-renormalizable
operators which violate the Lorentz symmetry. For the BDLV from 
the effective field theory, we discuss
the possible solutions to the above theoretical challenges, and we address the
Lorentz violation constraints from the LEP~\cite{bou1} and KamLAMD~\cite{:2008ee, review} 
experiments. For the BDLV from the Type IIB string theory
with D3-branes and D7-branes~\cite{Li:2011zm}, 
we point out that {\it the D3-branes are flavour blind, and
 all the SM particles are the conventional particles as in the traditional
SM when they do not interact with the D3-branes}. Thus,
we  not only can naturally avoid all the known
 phenomenological constraints on Lorentz violation, but also can naturally explain 
all the above theoretical 
challenges~\cite{Cohen:2011hx, GonzalezMestres:2011jc, Bi:2011nd, Cowsik:2011wv}. 
Moreover, we can explain the time delays for the high energy photons
compared to the low energy photons  
in  the MAGIC~\cite{Albert:2007qk}, HESS~\cite{Aharonian:2008kz}, 
and FERMI~\cite{Abdo:2009pg, Abdo:2009zza} experiments.
Such kind of models predicts that the photon
velocities linearly depend on their energies.
For a photon with energy  around a few GeV
we obtain that $\delta v_{\gamma}$ is around $10^{-5}$ on the Earth.
Thus,  we can test our models at the experiments
in principle.

\section{The Relevant Phenomenological Constraints and Observations on Lorentz Violation}

In this Section, we will briefly 
review the relevant phenomenological constraints and observations
on Lorentz violation.

First, the detection of neutrinos emitted from SN1987a gave us a lot of information 
not only on the process of supernova explosion, but also on neutrino properties. 
The Irvine-Michigan-Brookhaven (IMB)~\cite{imb}, Baksan~\cite{baks}, and Kamiokande 
II~\cite{kam} experiments collected $8+5+11$ neutrino events (presumably mainly $\bar\nu_e$)
with energies between $7.55~{\rm MeV}$ and $395~{\rm MeV}$ within 12.4 seconds. In particular,
the neutrinos arrived on the Earth about 4 hours before
the corresponding light. Although this  is compatible with the 
supernova explosion models, we can still obtain the upper limit on $\delta v_{\nu}$
\begin{eqnarray}
\left| \delta v_\nu (15~{\rm MeV})\right|   
 \le 2 \times 10^{-9}~.~\,
\label{spr2}
\end{eqnarray}
 This limit should be understood with an order-one uncertainty since
 the precise time delay between light and neutrinos is unknown.
Also,  we can employ the time coincidence of these events  to constrain 
the velocity differences for neutrinos with various energies. 
Rescaling a statistical analysis for the case with quadratic energy dependence,
 $\delta v_\nu \propto E^2$~\cite{Ellis:2008fc}, we obtain the bound
\begin{eqnarray}
\left| \delta v_\nu (30~{\rm MeV}) -\delta v_\nu (10~{\rm MeV}) \right|
 < 5 \times 10^{-13}\hbox{ at 95\% CL}~.~\,
\label{spr1}
\end{eqnarray}
However, there exists a larger uncertainty if the average neutrino energy changes 
with time during the detection interval around 10 seconds. 

Second, the MAGIC~\cite{Albert:2007qk}, HESS~\cite{Aharonian:2008kz}, 
and FERMI~\cite{Abdo:2009pg, Abdo:2009zza} Collaborations have
reported time-lags in the arrival times of high-energy photons, as compared with
photons of lower energies. The most conservative interpretations of
such time-lags are that they are due to emission mechanisms at the sources,
which are still largely unknown at present. However, such delays might also
 be the hints for the energy-dependent vacuum refractive index,
as first proposed fourteen years ago in Ref.~\cite{AmelinoCamelia:1997gz}.
Assuming that the refractive index $n$ depends linearly on the $\gamma$-ray
energy $E_{\gamma}$,
{\it i.e.}, $n_{\gamma} \sim 1 + E_{\gamma}/M_{\rm QG}$ where
$M_{\rm QG}$ is the effective quantum gravity scale, 
it was shown that the time delays observed by the MAGIC~\cite{Albert:2007qk}, HESS~\cite{Aharonian:2008kz}, 
and FERMI~\cite{Abdo:2009pg, Abdo:2009zza}  Collaborations are compatible with
each other for $M_{\rm QG}$ around $0.98\times 10^{18}$ GeV~\cite{Ellis:2009yx}.
 Also, there are the stringent
constraints coming from synchrotron radiation of the Crab
Nebula~\cite{crab,crab2,ems}. The D-particle models of space-time foam
have been proposed to explain all these effects
within the framework of string/brane theory,
based on a stringy analogue of the interaction of a photon with internal
degrees of freedom in a conventional medium~\cite{horizons,emnw,ems, Li:2009tt, Ellis:2009vq}. 
However, FERMI observation of GRB 090510 seems to allow only much
smaller value for time delay and then requires 
$M_{\rm QG} > 1.22 \times 10^{19}~{\rm GeV}$~\cite{:2009zq}. Because these data
probe different redshift ranges,  they may be compatible with each other
by considering a redshift dependent D-particle density~\cite{Ellis:2009vq}.

Third, the superluminal $\pi^+$  will lose energy quickly via the radiative emission
process $\pi^+\to\pi^+\gamma$. The IceCube experiment has measured the atmospheric
neutrino spectrum up to $\sim 400$ TeV, which agrees pretty well with the 
model calculations~\cite{Abbasi:2010ie}. Thus, the high energy charged 
pions do not lose much energy before they decay to neutrinos, which gives 
a strong constraint on the process $\pi^+\to\pi^+\gamma$. Note that the threshold of 
this process is $E_\pi > m_\pi/\sqrt{v_\pi^2-c^2}\simeq m_\pi/\sqrt{2 \delta v_\pi}$, and
the maximal pion energy is about $2$ PeV,
we obtain an upper bound on $\delta v_\pi < 2 \times 10^{-14}$~\cite{Abbasi:2010ie}.

Fourth, there are astrophysical constraints on Lorentz violation
for charged leptons, which are given as follows:

\begin{itemize}

\item The constraint from the Crab Nebula synchrotron 
radiation observations on the electron dispersion relation~\cite{crab2}.
For $\delta v_e$ linearly dependent on its energy, we obtain the
electron dispersion relation
\begin{eqnarray}
E^2 ~=~ p^2 +m_e^2 -  {{p^3} \over {M'_{\rm QG}}}~,~\,
\label{Elec-Disp}
\end{eqnarray}
where the corresponding effective quantum gravity scale $M'_{\rm QG} $ is 
larger than about $1 \times 10^{24}$ GeV~\cite{crab2}.

\item The electron vacuum Cherenkov radiation via the decay process $e\to e \gamma$ 
becomes kinematically allowed for $E_e >m_e/\sqrt{\delta v_e}$. Note that the cosmic ray electrons 
have been detected up to 2~TeV, we have $\delta v_e <  10^{-13}$ from cosmic ray experiments~\cite{bou2}. 

\item  The high-energy photons are absorbed by CMB photons and annihilate into the electron-positron pairs.
 The process $\gamma \gamma\to e^+e^-$ becomes kinematically possible for 
$E_{\rm CMB} > m_e^2/E_\gamma +\delta v_e E_\gamma/2$. Because it has been observed to occur 
for photons with energy about $E_\gamma =20$~TeV, we have
 $\delta v_e <2m_e^2/E_\gamma^2 \sim 10^{-15}$ from cosmic ray experiments~\cite{bou3} . 

\item The process, where a photon decays into $e^+e^-$,
 becomes kinematically allowed at energies $E_\gamma > m_e\sqrt{-2\delta v_e}$.  
As the photons have been observed 
up to 50 TeV, we have  $-\delta v_e < 2\times 10^{-16}$ from cosmic ray experiments~\cite{bou2}.
The analogous bound for the muon is $-\delta v_\mu < 10^{-11}$~\cite{Altmu}.

\end{itemize}

Fifth, there are the relevant constraints on Lorentz violation
from the experiments done on the Earth. Let us list them in
the following:

\begin{itemize}

\item The agreement between the observation and the theoretical expectation of 
electron synchrotron radiation as measured at LEP~\cite{bou1} gives the stringent
bound on isotropic Lorentz violation $|\delta v_e|< 5\times 10^{-15}$. Here,
we emphasize that the electrons and positrons propagated in the vacuum tunnel
at the LEP experiment.

\item For the electron neutrinos at the KamLAND experiment, the non-trivial
energy dependence of the neutrino survival probability implies
that the Lorentz violating  off-diagonal
elements of the $\delta v_{\nu}$ matrix in the flavor space
are smaller than about $10^{-20}$~\cite{:2008ee, review}.
Thus, if the Lorentz violation can not realize the flavour independent couplings
 naturally, we do need to fine-tune the relevant couplings.

\end{itemize}



\section{Background Dependent Lorentz Violation}

The main assumption for the background dependent Lorentz violation theories is:
{\it the Lorentz violation for all the SM particles is not constant
 in the space time}. In particular,
{\it the Lorentz violation for all the SM particles on the Earth is much larger
than those on the interstellar scale or in the vacuum}. For example,
to explain the results from the OPERA experiment and the SN1987a observations,
 we will show that the Lorentz violation on the
Earth is at least four orders larger than those on the
intestellar scale or in the vacuum.

We would like to discuss the background dependent Lorentz violation in
 a model independent way.
Considering the effective field theory or string theory, we can parametrize 
the generic $\delta v$ for a particle $\phi$ as follows
\begin{eqnarray}
\delta v_{\phi} ~=~ -{{m_{\phi}^2}\over {2 p^2}} + \sum_{n\ge 0} a^{\phi}_n {{p^n}\over {M^n_*}}~,~\,
\label{DV-Generic}
\end{eqnarray}
where $m_{\phi}$ and $p$ are respectively the mass and momentum of the particle, $a^{\phi}_n$
are the coefficients, and $M_*$ is the relevant effective scale. In the Type IIB 
string theory, we can obtain the $a^{\phi}_1$ term naturally by calculating the 
four-point function~\cite{Li:2011zm, Li:2009tt}, while in the effective 
field theory approach, 
we can realize all the terms in the above equation.
By the way, for a massless particle like the photon or  particle with tiny mass
like the neutrinos, we can neglect the mass term and change $p$ to $E$.

We denote the couplings $a^{\phi}_n$ on the Earth as $[a^{\phi}_n]^{\rm E}$, and
 the couplings $a^{\phi}_n$ on
the interstellar scale as $[a^{\phi}_n]^{\rm IS}$. As a concrete example, 
we employ the OPERA and SN1987a results to calculate the constraints on the ratios 
$[a^{\nu}_n]^{\rm IS}/[a^{\nu}_n]^{\rm E}$ for muon neutrinos. For simplicity,
we consider  $n=0,~1,~2$, and assume that only one $[a^{\nu}_n]^{\rm E}$ term generates
OPERA $\delta v_{\nu}$  in Eq.~(\ref{eq:opr}) and satisfies the SN1987a constraints
in Eqs.~(\ref{spr2}) and (\ref{spr1}). Thus, we obtain
\begin{eqnarray}
{{[a^{\nu}_0]^{\rm IS}}\over\displaystyle {[a^{\nu}_0]^{\rm E}}} \le 8.1 \times 10^{-5}~,~~~
{{[a^{\nu}_1]^{\rm IS}}\over\displaystyle {[a^{\nu}_1]^{\rm E}}} \le 2.3 \times 10^{-5}~,~~~
{{[a^{\nu}_2]^{\rm IS}}\over\displaystyle {[a^{\nu}_2]^{\rm E}}} \le 8.6 \times 10^{-3}~.~\,
\end{eqnarray}
Interestingly, considering the uncertainties in the SN1987a observations, we find that
 the ratio $[a^{\nu}_0]^{\rm IS}/[a^{\nu}_0]^{\rm E}$ is similar to the ratio
$[a^{\nu}_1]^{\rm IS}/[a^{\nu}_1]^{\rm E}$. Notice that both the $[a^{\nu}_0]^{\rm E}$ term 
and the $[a^{\nu}_1]^{\rm E}$ term
can be consistent with the OPERA results on weak energy dependence
for $\delta v_{\nu}$ or $\delta t$, thus,
these results may be generated by both terms. In particular, we propose that
the OPERA experiment can test the $\delta v_{\nu}$ energy dependence  
by lowering their muon neutrino energies to about a few GeV. If the 
$\delta v_{\nu}$ energy dependence is confirmed, it will be a big discovery for sure.

In addition, the speed of light in the vacuum has been measured on the Earth. We 
conjecture that there may exist a ``Dream Special Relativity Theory''. In particular,
the photon velocity on the Earth is not the maximal photon velocity in the 
``Dream Vacuum'' due to the background effects. 
Thus,  in principle, {\it all the SM particles including
both the photons and neutrinos are subluminal 
after Lorentz violation}. We can explain the OPERA and MINOS results as well. 
As a concrete example, we can
consider that all the SM particles have the same $a^{\phi}_0$, $a^{\phi}_1$,
and $a^{\phi}_2$, and we require 
 $a^{\phi}_0 < 0 $ and $a^{\phi}_2 < 0 $ while $a^{\phi}_1 > 0$. The point is that
the speed of light has been measured on the Earth 
for the photons with very low energies.
The more detailed study will be given
elsewhere. Interestingly, although we can not measure the
$[a^{\gamma}_0]^{\rm E}$ term in the Laboratories on the Earth, we can try to measure
the $[a^{\gamma}_n]^{\rm E}$ terms in the  Laboratories on the Earth for $n \ge 1$ by varying
the photon energies if the photon energies are not very small.

Let us subscribe to the Eddington's dictum:
  ``Never believe an experiment until it has been confirmed by theory''.
Thus, we need to understand
 the $a_n^{\nu}$ terms in Eq.~(\ref{DV-Generic}) from the theoretical
point of view. First, we consider the effective field theory, and will
 generate  the $a_n^{\nu}$ terms for $n=0,~1,~2$ simultaneously, which
has not been studied yet.  As we know,
to explain the neutrino masses and mixings in the Type II seesaw 
mechanism, we need to introduce
a triplet Higgs field $\Phi$ whose quantum number under 
$SU(3)_C\times SU(2)_L \times U(1)_Y$ gauge symmetries
is $(\mathbf{1}, \mathbf{3}, \mathbf{1})$~\cite{GRmodel, triplet1}.
To break the Lorentz symmetry, we choose a unit vector $u^{\mu}=(1,~0, ~0, ~0)$,
which does not break the  $SO(3)$ space rotation symmetry. 
This can be realized by giving Vacuum Expectation Value (VEV)
to a vector field or $\partial_{\mu} \phi$.
Therefore, we can have  the following non-renormalizable terms in
the Lagrangian
\begin{eqnarray}
{\cal L} &=& {1\over {1+\delta_{ij}}} \left( 
i y_{jk} {{ \Phi} \over {M_*}} u^{\mu} L_j \partial_{\mu}  L_k
+ y'_{jk} {{ \Phi} \over {M^2_*}} (u^{\mu} \partial_{\mu}  L_j)
(u^{\nu} \partial_{\nu}  L_k) \right)
~,~\,
\end{eqnarray}
where $y_{jk}$ and $y'_{jk}$ are coupling constants,
and $L_i$ denote the lepton doublets.
For simplicity, we assume  $y_{jk}=0$ and $y'_{jk}=0$ for
$j\not= k$. Thus, we obtain 
\begin{eqnarray}
\delta v_{\nu} &=& {{y_{ii}^2 V^2_{\Phi}}\over {2 M_*^2}}
+{{2y_{ii} y'_{ii} V^2_{\Phi} E}\over { M_*^3}} 
+  {{3 y^{\prime2}_{ii} V^2_{\Phi} E^2}\over {2 M_*^4}} ~,~\,
\end{eqnarray}
where $V_{\Phi}$ is the VEV of $\Phi$.
In short, we can indeed have the $a_n^{\nu}$ terms for $n=0,~1,~2$
in the mean time from the effective field theory. 
Similar discussions hold
for the  $a_n^{\nu}$ terms with $n > 2$.

Furthermore, the  $a_1^{\phi}$ terms in Eq.~(\ref{DV-Generic}) 
can be generated in the Type IIB string theory~\cite{Li:2011zm, Li:2009tt}.  
Let us briefly review the results.
We consider the Type IIB string theory with D3-branes
and D7-branes where the D3-branes are inside the D7-branes~\cite{Li:2011zm, Li:2009tt}.
The D3-branes wrap a three-cycle, and the D7-branes wrap a four-cycle.
Thus, the D3-branes can be considered as point
particles in the Universe, while
the SM particles are on the world-volume of the D7-branes.
We assume that the $V_{A3}$ is the average three-dimensional
volume which has a D3-brane locally in the Minkowski space dimensions, and $R'$ is the
radius for the fourth space dimension transverse to the D3-branes in the D7-branes.
Especially, $V_{A3}$ is the inverse of the D3-brane number density and can vary
in the space time. Also, we define 
the dimensionless parameters $\eta$ and $\xi$ as follows~\cite{Li:2011zm}
\begin{eqnarray}
\eta \equiv {{(1.55)^4} \over {V_{A3} R'M^4_{\rm St}}}~,~~~
\xi \equiv M_{\rm St} \times V^{1/3}_{A3}~,~\,
\label{ETAXI-P}
\end{eqnarray}
where $M_{\rm St}$ is the string scale. Especailly,
$\eta$ is a small number around 0.1 or smaller for a perturbative theory.

For the $SU(3)_C \times SU(2)_L \times U(1)_Y$ gauge fields and their corresponding 
gauginos, we have~\cite{Li:2011zm} 
\begin{eqnarray}
\delta v \simeq \pm {{(2n+1)\pi E}\over\displaystyle {\xi M_{\rm St} }}~.~\,
\label{GF-G}
\end{eqnarray}
While for all the other SM particles, we have~\cite{Li:2011zm}
\begin{eqnarray}
\delta v \simeq \pm {{(2n+1)\pi \eta E}\over\displaystyle {\xi M_{\rm St} }}~.~\,
\label{OSMP}
\end{eqnarray}
Let us consider the lowest order term $n=1$. As shown in Ref.~\cite{Li:2011zm},
we can explain the MAGIC, HESS, FERMI, OPERA, and MINOS experiments 
simultaneously. Moreover, if we do not consider the time delays in
the MAGIC, HESS, and FERMI experiments, we can choose the positive
signs in both Eq.~(\ref{GF-G}) and Eq.~(\ref{OSMP}).
Note that the speed of light has been measured for the very low energy
photons,  we can still explain the OPERA results, which is similar to 
the discussions in Ref.~\cite{Li:2011zm}. The point is that 
 $\delta v_{\gamma}$ for a very low energy photon can be much smaller than
$\delta v_{\nu}$ for a muon neutrino with energy 17~GeV.
Interestingly, note that $\delta v_{\nu}/\delta v_{\gamma} = \eta$ if the
photon and neutrino have the same energies,
we obtain that for a photon with energy  around a few GeV
 $\delta v_{\gamma}$ can be around $10^{-5}$ on the Earth.
Thus,  we would like to propose new experiments to determine
whether the photon velocities are dependent on their energies or not.

\section{The Solutions to the Theoretical Challenges of the OPERA Experiment}

With background dependent Lorentz violation, we  assume that the Lorentz violation
on the Earth is much larger than those 
on the interstellar scale. So we can automatically escape the
stringent astrophysical constraints on Lorentz violation. However, we still need to 
address the constraints on Lorentz violation from the LEP and KamLAND
experiments. In particular, there are two theoretical challenges to the OPERA experiment:
the  Bremsstrahlung effects for muon neutrinos~\cite{Cohen:2011hx} 
and the pion decays~\cite{GonzalezMestres:2011jc, Bi:2011nd, Cowsik:2011wv}.
Thus, we will study these issues in background dependent Lorentz violation
scenarios from the effective field theory and Type IIB string theory.

\subsection{Background Dependent Lorentz Violation from the Effective Field Theory}

To explain the OPERA results and the astrophysical constraints/observations
on Lorentz violation, we must 
assume that the relevant effective operators for Lorentz violation are confined to
the Earth. For example,  the triplet Higgs field
 $\Phi$ in our model must be on the Earth.

To avoid the constraints from the KamLAND experiment, we do need to fine-tune
the relevant couplings since the Lorentz violation is not flavour blind in general. 
Moreover, the constraint on $\delta v_{\pi}$ arises from the astrophysical
observations, and then
$\delta v_{\pi}$ can be comparable to $\delta v_{\nu}$ on the Earth.
Thus, the pion decays are not a problem. The Bremsstrahlung effects 
for muon neutrinos and the LEP constraint on $\delta v_e$ are subtle.
If the vacuum in the LEP tunnel is similar to the vacuum on the interstellar
scale,  we can indeed explain them simultaneously: 
$\delta v_e$  can be  smaller than $5\times 10^{-15}$  in 
the LEP tunnel, while it can be very close to $\delta v_{\nu}$ on the Earth.
Thus, the threshold energy of muon neutrinos for the process 
 $\nu_{\mu} \to \nu_{\mu} + e^+ + e^-$ can be larger than 100 GeV or even higher. 

However, if $\delta v_e$ in the LEP tunnel is similar to that on
the Earth, we may have a problem here. One possible solution is that
 we  consider the scenarios where the Lorentz symmetry is deformed.
The other possible solution, which we can imagine, is that
both photons and neutrinos are subluminal. In short, the
detail calculations on  the Bremsstrahlung effects 
for muon neutrinos definitely deserve further careful study.

\subsection{Background Dependent Lorentz Violation from the Type IIB String
Theory}

For the background dependent Lorentz violation in the Type IIB string
theory with D3-branes and D7-branes, we can simultaneously
explain the OPERA and MINOS results, avoid all the astrophysical constraints
on Lorentz violation,
and explain the time delays in
the MAGIC~\cite{Albert:2007qk}, HESS~\cite{Aharonian:2008kz}, 
and FERMI~\cite{Abdo:2009pg, Abdo:2009zza} experiments 
simultaneously~\cite{Li:2011zm}. 

Because our D3-branes are flavour blind, we  naturally explain 
the KamLAMD experiments.   The time delays or advances
 for the SM particles arise from
their interactions with the D3-branes. Thus, when they do not
interact with the D3-branes, for example, when they are away from the D3-branes, 
all the SM particles are just the conventional particles in the traditional 
SM without any Lorentz violation at all. Thus, we not only naturally
avoid the LEP constraint on $\delta v_e$, but also naturally
explain the theoretical challenges
such as the Bremsstrahlung effects for muon neutrinos and the
pion decays. In fact, all the astrophysical Lorentz violation constraints
on charged leptons 
are automatically escaped by the same reason.

The only constraints on our models arise from the time delays in the 
 MAGIC, HESS, and FERMI experiments and the Lorentz violation constraint 
from the SN1987a observations. For a photon with energy around 
a few GeV, we obtain $\delta v_{\gamma} \sim 10^{-5}$ on the Earth.
Note that the photon velocities are linearly dependent on
their energies in our model, we can test such proposal by doing experiment
to measure the velocities of the photons with different energies.
We shall study it further in the future.



\section{ Conclusions}

To explain both the OPERA experiment and all the known phenomenological  
constraints/observations on Lorentz violation, 
the background dependent Lorentz violation  has
been proposed. We studied the BDLV in a  model independent way,
and conjectured that  there may exist a ``Dream Special Relativity 
Theory'', where all the SM particles can be subluminal 
due to the background effects.
We also suggested that the OPERA experiment can measure the 
velocities of the muon neutrinos with energies around a few GeV, which
may test whether the muon neutrino velocities depend on their energies or not.
Assuming that the Lorentz violation on
the Earth is much larger than those on the interstellar scale, we automatically
escaped all the astrophysical constraints on Lorentz violation.
For the BDLV from the effective field theory,  we considered a simple model 
with a triplet Higgs field as in the Type II seesaw mechanism, 
where we introduced the non-renormalizable operators which violate 
the Lorentz symmetry. We discussed the possible solutions to the 
theoretical challenges 
such as the Bremsstrahlung effects for muon neutrinos and the
pion decays. Also, we addressed the
Lorentz violation constraints  from the LEP and KamLAMD 
experiments. For the BDLV from the Type IIB string theory with D3-branes and D7-branes,
 we point out that the D3-branes are flavour blind, and
 all the SM particles are the conventional particles as in the traditional
SM when they do not interact with the D3-branes. Thus,
we  not only can naturally avoid all the known
 phenomenological constraints on Lorentz violation, but also can naturally explain 
all the theoretical challenges. Moreover, we can explain the time delays for 
the high energy photons compared to the low energy photons in the MAGIC, HESS, and FERMI
experiments simultaneously. 
It was predicted that the photon
velocities linearly depend on their energies.
For a photon with energy  around a few GeV
we obtained that $\delta v_{\gamma}$ is around $10^{-5}$ on the Earth.
Thus,  we can test our models at the experiments
in principle.

\section*{Acknowledgment}

This research was supported in part 
by the Natural Science Foundation of China 
under grant numbers 10821504 and 11075194 (TL),
and by the DOE grant DE-FG03-95-Er-40917 (TL and DVN).


\end{document}